\documentstyle[12pt,prl,aps,preprint,tighten,floats]{revtex}
\begin{document}

\title{ Magnetic Black Holes in String Theory}
\author {Mikhail Z. Iofa\\
Skobeltsyn Institute of Nuclear Physics \\
Moscow State University\\
Moscow 119899, Russia\\ }

\def\f0{f_{0}}
\def\f0{f_{0}}
\def\pp{\partial_{z}}
\def\ppm{\partial_{\bar{z}}}
\def\pd{\partial_{\mu}}
\def\pa{\partial}
\def\pu{\partial^{\mu}}
\def\P{\Phi}
\def\p{\phi}
\def\L{\Lambda}
\def\S{\Sigma}
\def\l{\lambda}
\def\m{\mu}
\def\n{\nu}
\def\e{\epsilon} 
\def\s{\sigma}
\def\te{\theta}
\def\g{\gamma}
\def\d{\delta}
\def\D{\Delta}
\def\a{\alpha}
\def\b{\beta}
\def\r{\rho}
\def\vp{\varphi}
\def\hP{\hat P}
\def\F{{\cal F}}
\maketitle

{\large\bf 1}. At present, string theory is considered the best candidate for
 a fundamental theory that would provide a consistent quantum theory of
gravity unified with the other interactions .
In particular, string theory provides a powerful approach to the physics of
black holes. In this setting a fundamental problem is that of
understanding how the intrinsically stringy effects modify  the Einstein
gravity. In this paper we discuss two of these effects:  presence of
scalar fields such as the dilaton and moduli and higher-genus
contributions modifying  the tree-level effective action. 
We focus on higher-genus corrections, because 
string theory, being a theory formulated on a world sheet, always contains
string-loop corrections from higher topologies of the world sheet (these
vanish only for higher supersymmetries $N\geq 4$), while ${\a}'$ corrections 
can vanish  in  certain constructions based on conformal field theories
\cite{howe,bh2} and 
for a large class of backgrounds \cite{bh2,bh1,cvyo}.  

We consider the perturbative 
 one-string-loop (torus topology) corrections for a special class of 
backgrounds: $4D$ black holes as solutions to the equations of motion of the 
4D effective field theory provided by  heterotic string theory 
compactified on the manifold $K3\times T_2$ or on a suitable orbifold.
For this compactification pattern, the effective $4D$ field theory is
locally $N=2$ supersymmetric theory \cite{duff}. String-loop
corrections modify the tree-level effective action which, due to $N=2$
supersymmetry, receives only one-string-loop contributions
\cite{uni,kou,wikalu,defekoz}.

Taking the string-loop expansion parameter $\e$ to be small, we can make the
perturbative expansion of the equations of motion and
obtain a perturbative solution of the loop-corrected equations of motion.

In the case of magnetic black hole,  the dilaton of the string-tree-level 
solution $\p =\ln\frac{(r+P_{1} )(r+P_{2})}{r^2} $ increases and the
tree-level
gauge coupling $e^{-\p}$ decreases at small distances, so that the effective 
loop-corrected gauge coupling is sensitive to the string-loop correction. 

The universal sector contains supergravity and vector multiplets, the vector
components resulting from dimensional reduction of $6D$ supergravity on the
two-torus $T_2$.
We calculate the effective gauge couplings in the universal sector by using
the (perturbative) loop-corrected prepotential. Due to the $N=2$ 
supersymmetry, there are no loop corrections
to the prepotential beyond  one loop \cite{wikalu,hm}.

Having the string-loop-corrected effective action, we derive the equations
of motion and solve them analytically in the first order in the
loop-counting parameter $\e$. Technical simplification is achieved in the case
of equal magnetic charges, in which case the tree-level moduli and loop 
corrections taken with the tree-level moduli are constants. 
However, the results remain unchanged
in the general case $P_1 \neq P_2$.

In the first order in $\e$, we obtain the loop-corrected metric
\begin{equation}
\label{F0}
g^{ii}= -g_{00} =\left(1+\frac{P}{r}\right)\left(1 + \e A
\left(\frac{P}{r}-\frac{P}{r+P}\right)\right). 
\end{equation}
Supersymmetric solution is obtained for $A=\frac{V}{4}$,
where the Green-Schwarz function $V$ is strictly positive \cite{kou}.
At small $r$, up to the terms of order $O(\e^2)$, the 
metric (\ref{F0}) with  $A=\frac{V}{4}$ can be rewritten in the form
$$-g_{00} = 1+\frac{P}{r+\e\frac{V}{4}},$$ 
where the singularity at the origin is smeared by quantum corrections. 

{\large\bf 2.} An  effective field theory which describes dynamics of  light
fields in four dimensions depends on  the pattern of compactification of
the initial superstring theory. 
Compactification of heterotic string theory on the manifold $K3\times T^2$
or on a suitable orbifold with the $SU(2)$ holonomy 
yields 4D theory with N=2 supersymmetry.

The bosonic part of the universal sector of the
 the effective field theory which includes gravity, dilaton and the
antisymmetric tensor is
\begin{equation}
\label{F1}
I_6=\frac{1}{2{\kappa_6}^2}\int d^6 x\sqrt{-G}e^{-{\Phi}'}\left 
[R + (\partial {\Phi}'
)^2 - \frac{H^2}{12} \right ]+\ldots.
\end{equation}

Compactification of the $6D$ theory on two-torus yields the action
\cite{sen} (we retain only non-vanishing backgrounds relevant for the
following)
\begin{eqnarray}
\label{E7}
S&=&\frac{1}{\kappa^2}\int d^4 x\sqrt{-G'}e^{-{\p}'}
\left[R + (\partial{\p}')^2 -\frac{(H')^2}{12} +
\frac{1}{4}\partial G_{11}\partial G^{11} +
\frac{1}{4}\partial G_{22}\partial G^{22} +...
\right. \nonumber \\
&-&\left. \frac{1}{4}G_{11} (F^{(1)1})^2-\frac{1}{4}G_{22} (F^{(1)2})^2
 - \frac{1}{4}G^{11} (F^{(2)}_1)^2
-\frac{1}{4}G^{22} (F^{(2)}_2)^2 \right] +\ldots.
\end{eqnarray}

Here 
$$G'_{\mu\nu} = G_{\mu\nu} - G_{mn} A^{(1)m}_{\mu} A^{(1)n}_{\nu}, \quad
H'_{\m\n\l} = H_{\m\n\l} -A^{(1)n}_{[\m}H_{\n\l ] n} +A^{(1)m}_{[\m} 
A^{(1)n}_\n H_{\l ]mn} 
$$
$$ \p' =\Phi' -\frac{1}{2}\ln\det (G_{mn}). $$
 $G_{mn}\,\,\,( m,n =1,2)$ are the components of the internal metric,
$B_{mn}=0$ and the field strengths are obtained from the potentials of the
form $A^{(1)n}_{\mu}  =  G^{nm} G_{m\mu}$ and $A^{(2)}_{n\mu } =  B_{n\mu}$.
Magnetic field potentials are
\begin{eqnarray}
A^{(1)1}_{\mu}  =(0,\ a_s),\qquad 
A^{(2)}_{1\mu }  =(0,\ b_s).
 \label{E2}   
\end{eqnarray}

Purely electric and magnetic solutions can be embedded in
$4D$ N=2 supersymmetric theory and leave 1/2 of the supersymmetry unbroken
\cite{cvyo}.
We shall consider the
following solution of the equations of motion derived from (\ref{E7})
\cite{bh2,cvyo}:
\begin{eqnarray}
\label{E9}
ds^2=-{\f0}^{-1}(r)dt^2 +{\f0}(dr^2+r^2d\Omega_2^2),\quad \f0 =
1+\frac{P}{r}, \nonumber \\
\quad \p' =\ln \f0, \quad  
a_\varphi  =a^{-1}P(1-\cos\vartheta),\quad b_\varphi =aP(1-\cos\vartheta),
\end{eqnarray} 
where $a_\varphi$ and $b_\varphi$ are nonzero components of potentials in
spherical coordinates, $a=Const$.

Tree-level components of the internal metric $G_{mn}$ and  magnetic
field strengths are
\begin{eqnarray}
\label{B1}
G_{11} =a^2 , \qquad \,\, G_{22} =A^2 ,\nonumber \\
F^{1(1)}_{ij} =a^{-1}F_{ij}, \quad F^{(2)}_{1ij} =aF_{ij}, \quad F_{ij} =
-\varepsilon_{ijk}\pa^k \f0,
\end {eqnarray}
 and in spherical coordinates have a single
 nonzero component $F_{\vartheta\varphi}= P\sin\vartheta $.
The fields $H'_{\m\n\l}$ for the solution in question vanish.

{\large\bf 3.} Considered within the framework of $N=2$ locally supersymmetric theory
\cite{dewit}, the action
(\ref{E7}) can be identified with the bosonic part of the universal 
sector of the complete action.  The form of the $N=2$ supergravity depends 
on the choice of a holomorphic
vector bundle $(X^A , F_A )$, where $X^A$ denote both the vector superfields
and the complex scalar components of these superfields. Different
sections correspond to equivalent equations of motion and are connected by
symplectic transformations \cite{wikalu,dewit,cafvp}. If a holomorphic
bundle admits a holomorphic prepotential
$F(X)$, the latter completely defines  dynamics of  $N=2$ supersymmetric
theory.
In particular, in this case $F_A = \pa F/{\pa X^A }$. 

The dilaton
${\Phi}'$ in (\ref{F1}) can be split in the sum ${\Phi} + {\Phi}_\infty $
where ${\Phi}$ can be chosen so that
at the spatial infinity ${\Phi}=0$ and ${\Phi}_\infty $ is a free
parameter.
We split also $ {\p}' ={\p} + {\p}_\infty $, where
${\Phi}_\infty = {\p}_\infty $. The loop counting parameter is $\e
=e^{{\p}_\infty}$.

 It is proved that the prepotential has no perturbative
loop corrections beyond one loop and is of the form \cite{wikalu,defekoz,hm}
\begin{equation}
\label{M6}
F=-\frac{X^1 X^2 X^3}{X^0} -i {X^0}^2 h^{(1)}(-i\frac{X^2}{X^0},
-i\frac{X^3}{X^0}).
\end{equation}

We use the
standard identification of special coordinates (for real moduli, i.e. for
vanishing axion, diagonal metric $G_{mn}$ and $B_{mn}=0$)
$$ \frac{X^1}{X^0}  =iS, \quad \frac{X^2}{X^0}=iT=i\sqrt{G_{11}G_{22}}, \quad 
\frac{X^3}{X^0}=iU=i\sqrt{G_{11}/G_{22}}, $$
and
$$ S=e^{\p},\quad T\equiv e^{\g +\s},\quad U\equiv e^{\g -\s}.$$
In perturbative approach, we neglect (small) non-perturbative corrections
 to the prepotential of the form $f(e^{-2\pi S}, T,U)$. 
From (\ref{M6}) it follows that the loop corrections are
multiplied by a  factor  $\e $ and can be treated perturbatively.
The loop correction $h^{(1)}$ is
independent of $S$,  
 loop corrections to the tree-level expressions 
are everywhere multiplied by the factor $\e$.
  
The kinetic terms for the moduli can be written in terms
of the Kaehler potential $K=-\log i(\bar{X}^A F_A -{X}^A \bar{F}_A)$. 

The loop-corrected Kaehler potential \cite{kou,wikalu,defekoz,afgnt}  is
calculated by using the loop-corrected prepotential and is of the
form
\begin{equation}
\label{M7}
K = -\log[(T+\bar{T})(U+\bar{U})(S+\bar{S}
+\e V(T,\bar{T},U,\bar{U})],
\end{equation}
where the Green-Schwarz function $V$ written for real function $h^{(1)}$ and
real moduli $T$ and $U$ is 
\begin{equation}
\label{M8}
V(T, U)=\frac{v(T,U)}{T\,U} = \frac{h^{(1)} -T \pa_{T} h^{(1)}-U \pa_{U} h^{(1)}}
{T\,U}.
\end{equation}

The kinetic terms of gauge fields are
$$ -\frac{i}{8} \left( N_{IJ} F^{+I}_{\mu \nu }F^{+\mu \nu J} -\bar{N}_{IJ}
F^{-I}_{\mu\nu }F^{-\mu\nu J}\right),$$
where $F^{\pm }$ are the self-dual and anti-self-dual field strengths.
The couplings $N_{IJ}$ are \cite{dewit}
\begin{equation}
\label{M9}
N_{IJ}=\bar{F}_{IJ} +2i\frac{(Im F_{IK} X^K )(Im F_{JL} X^L )}
{(X^I Im F_{IJ} X^J )}, 
\end{equation}
where $F_{IJ} ={\pa}^2 F/{\pa X^I \pa X^J} $ and $F$ is the total
loop-corrected prepotential.

Starting from heterotic string theory, one naturally arrives at the
effective action written in terms of the  "heterotic"
 symplectic vector bundle. 
In this basis, the moduli are treated non-symmetrically, the modulus
$S$ playing the role of the coupling constant \cite{wikalu,defekoz}, 
and the vector
couplings become weak in the large-dilaton limit.
It is known that in terms of the heterotic symplectic bundle it is
impossible to introduce a prepotential, however, one can construct the
 prepotential in another sections in which the prepotential exists and
calculate in this basis the loop correction to the prepotential
\cite{wikalu,cafvp}.  Following \cite{wikalu,cafvp}, we introduce a new basis 
$(\hat{X}, \hat{F})$ which admits the prepotential by performing the symplectic 
transformation
$$
\left(\begin{array}{c}
\hat{X}\\ \hat{F}\end{array}\right) =O\left(\begin{array}{c}
X\\ F\end{array}\right)\qquad O=
\left(\begin{array}{cc}
U &Z\\W &V \end{array}\right),  
$$
where the nonzero elements of the matrix $O$ are $U^I_{\,\,J}
=V_{J}^{\,\,I}=\delta^I_J$ for $I,J \neq 1$, $W_{11} =-Z^{11}=\pm 1$.
The gauge coupling constants transform as \cite{wikalu,cafvp}
\begin{equation}
\label{B7}
 \hat{N} = (VN+W)(U+ZN)^{-1}.
\end {equation}
The Kaehler potential is invariant under the symplectic transformations.

The loop-corrected
couplings are calculated using (\ref{M9}) and the full prepotential
(\ref{M6}). In  $N=2$ locally supersymmetric theory the Wilsonian gauge
couplings which
enter the effective action receive only one-string-loop corrections
\cite{uni,wikalu,defekoz,ant}.             	  

 By using (\ref{M9}) we calculate  the gauge couplings  
in the basis which admits the prepotential  and using (\ref{B7}) we 
transform them to the "heterotic" basis. We obtain the gauge part of the
loop-corrected action in the form
\begin{eqnarray}
\label{B4}
L_g &=& -{1\over 4}\left[\left(e^{-\p +2\g } -\e\frac{n}{4}\right) (F^{(1)1})^2 +
\left(e^{-\p-2\g } -\e\frac{n}{4}e^{-4\g}\right) (F_1^{(2)})^2\right.
\nonumber \\ 
&+&\left. \e\frac{n+2v}{4}e^{-2\g}(F^{(1)1}F_1^{(2)})\right],
\end {eqnarray}
where 
$ n=v(T,U) +T^2 \pa_T^2 h^{(1)} +2TU\pa_T\pa_U h^{(1)} +U^2 \pa_U^2 h^{(1)}. $

In the case of the standard embedding of the spin connection in the gauge
group $E_8\times E_8$ which leaves the unbroken group $[E_8\times E_7\times
U(1)^2]_L \times [U(1)^2]_R$ (or its enhancement of at
special points in the moduli space) and
in the $Z_2$ orbifold limit, the loop-corrected prepotential was calculated
in \cite{hm}, however its specific properties are not relevant here.

{\large\bf 4.} Using the loop-corrected expressions for kinetic terms for the moduli and
the gauge part of the action and
retaining only two magnetic field strengths, we obtain 
the loop-corrected $4D$ action (in the Einstein frame) as (cf. (\ref{E7}):
\begin{eqnarray}
\label{E10}
S&=&\frac{1}{\kappa^2}\int d^4 x\sqrt{-g}
\left[R-\frac{1}{2}(\pa\p)^2(1-\e e^{\p}V)-(\pa\s)^2(1+ \e e^\p F_{\s\s})
 \right.\nonumber \\
&-& \left. (\pa\g)^2(1+ \e e^\p F_{\g\g})
-2\pa\s\pa\g \e e^\p F_{\s\g} - 2\pa \p 
\pa\s \e e^\p F_{\p\s} -2\pa\p \pa\g \e e^\p F_{\p\g}
\right. \nonumber \\
&-&
{1\over 4}\left[\left(e^{-\p +2\g } -\e\frac{n}{4}\right)
(F^{(1)1})^2 +
\left(e^{-\p-2\g } -\e\frac{n}{4}e^{-4\g}\right) (F_1^{(2)})^2 \right.
\nonumber \\ 
&+& \left.\e\frac{n+2v}{4}(F^{(1)1}F_1^{(2)})\right],
\end{eqnarray}
where
$
F_{\p\s} =-\frac{1}{4} (V_T +V_U +V_{\bar{T}} + V_{\bar{U}}), \quad
F_{\s\g} =- (V_{T\bar{T}} +V_{U\bar{U}}), \ldots .
$
Here the indices at $V$ denote the derivatives with respect to the
corresponding variables.

We look for a static spherically-symmetric solution of the field equations 
of the loop-corrected action (\ref{E10}).
The general ansatz for the metric is 
\begin{equation}
\label{E12}
ds_4^2 =- e^{\nu} dt^2 + e^{\l}dr^2 + e^{\mu}d\Omega_2^2.
\end{equation}
The field strengths $F^{(1)}_{ij}$ and $F^{(2)}_{ij}$ are  assumed to have 
the same functional form as
at the tree level
\begin{equation}
\label{D3}
 F^{(1)}= e^{2\g_0 }F, \quad  F^{(2)}= e^{-2\g_0 }F, \quad
F_{\vartheta\varphi}= P\sin\vartheta . 
\end{equation}

In the first order in  $\e$, we look for a solution in the form
\begin{eqnarray}
\nu =-\ln\f0 +\e n ,\,\, \l =\ln\f0 +\e l ,\,\, \mu =\ln\f0 +2\ln r +\e m,
 \,\, \p = \ln\f0 + \e\varphi
\label{E25}
\end{eqnarray}
Here $n,m,l,\p$ and $f$ are unknown functions which are determined from the 
field equations.

At the tree level, the moduli $\s$ and $\g$ are constants. Thus, we can
 write $\pd\s =\e\pd\s_1 ,\,\, \pd\g =\e\pd\g_1 $. 
All the kinetic terms in (\ref{E10}) which contain $\s_1$ or $\g_1$
are of order $O(\e^2)$. 

The equation of motion for the dilaton is
\begin{eqnarray}
\label{B5}
\frac{1}{\sqrt{-g}}\pa_{\mu }\left(g^{\mu\nu }\sqrt{-g}(\pa_{\nu }
\p (1-\e V e^\p )\right)+\frac{1}{2} (\pa\p)^2 \e Ve^\p+ \nonumber \\
\frac{1}{4} \left(e^{-\p+2\g} (F^{(1)1})^2 +
e^{-\p-2\g}(F_1^{(2)})^2\right)=0.
\end{eqnarray}
All the terms containing the functions
$F_{\s\s}, F_{\s\g}, F_{\s\p}, F_{\g\p}$ and $ F_{\g\g}$ can be dropped as
they are of the next order on $\e$, and the terms with $\g_1$ in the gauge terms 
 cancel. We are left with
\begin{eqnarray}
\label{E14}
\frac{1}{\sqrt{-g}}\pa_{\mu }\left(g^{\mu\nu }\sqrt{-g}\pa_{\nu }
\p \right)
-\frac{1}{2} \e e^{\p} (\pa\p)^2 V 
 + \frac{1}{2} e^{-\p}F^2 (1+\e e^{\p} V) =0.
\end{eqnarray}
 Here and below it is understood that all the expressions containing the 
factor $\e$ are calculated with the tree-level fields.

The field equations and the Bianchi identities for the gauge field strengths
are
\begin{equation}
\label{L3}
\pd \left(\sqrt{-g} Im N_{IJ} \F^J + Re N_{IJ} \F^{*J}\right)^{\m\n} =0
\end{equation}
and
\begin{equation}
\label{L4}
\pd \F^{*J\m\n} =0,
\end{equation}
where we used notations adapted to notations of the $N=2$ supersymmetric 
formulation
$$ F^{(1)1}=\F^0, \qquad F_1^{(2)}=\F^1,  $$ 
and $\F^{*J}_{\m\n}= \frac{1}{2}\sqrt{-g}\e_{\m\n\rho\l}\F^{J\rho\l}$.
  
At the tree level $\F^{0,1\, 0r} =0$, and (\ref{L3}) is satisfied
identically. At the one-loop level, noting that only   $ N_{00}$ and  $
N_{11}$ are 
$O(1)$ and $ N_{0J}$ and  $ N_{1J}$  are $O(\e )$, we
obtain solution of (\ref{L3}) in the form
$$ \F^{I\, or} =\frac{\e c_I (\vartheta,\varphi)}{\sqrt{-g'} Im
N_{II}},\qquad I=0,1, $$
where $-g'(r) =e^{\n +\l +2\m}$ and $c_I (\vartheta,\varphi)$ are
arbitrary functions. 
The Bianchi identity (\ref{L4}) shows that $c_I = Const$. 

The $\n =\vartheta $ components of Eq.(\ref{L3}) with $I=0,1$ are (the
$\varphi$ components yield the same result)
$$\pa_{\vartheta}  \left(\sqrt{-g} Im N_{I0} \F^0 + Im N_{I1}
\F^{1}\right)^{\vartheta\varphi} =0 $$  
and for our ansatz are satisfied identically. The Bianchi identities are $$
\pa_r
\F_{\vartheta\varphi}^{0,1}
=0  $$ yielding the field strengths in the form
$$\F_{\varphi\vartheta}^{0,1} =P^{0,1}\sin \vartheta.$$

With the required accuracy, the Einstein equations can be written as
\begin{equation}
\label{E16}
R_{\mu\nu}-\frac{1}{2}g_{\mu\nu}R -\frac{1}{2} 
\left(\pd\p \pa_{\nu}\p -\frac{1}{2}g_{\mu\nu}(\pa\p)^2 \right)\left(1-\e V
e^\p \right)+ {(L_g)}_{\m\n} -{1\over 2} g_{\m\n}L_g =0
\end{equation}
Here
\begin{equation}
 L_g = {1\over 4}\left[2e^{-\p} +\e V \right] F^2, \quad
{(L_g)}_{\m\n}={1\over 4}\left[2e^{-\p} +\e V \right]
(F^2)_{\m\n}. 
\end{equation}
The functions $\s_1$ and $\g_1$ decouple from the above equations.
Since in this paper we are primarily interested in the form of loop-corrected 4D metric, 
there is no need to determine these functions explicitly.

{\large\bf 5.} Next we substitute the functions $\m, \n, \l$ and $\p$ in 
the form (\ref{E25}) in the above equations.
In the first order in parameter $\e$ the Einstein equations (with one index
lifted) are 
\begin{eqnarray}
\label{E22}
m''+m'(q'+\frac{3}{r})-l'(\frac{1}{2}q'+\frac{1}{r})
+{\vp}'\frac{q'}{2} -\frac{l-m}{r^2}+\frac{1}{2}{q'}^2 s  =0,\nonumber \\
m''+n'' +m'\frac{2}{r} -l'\frac{1}{r} + n'(-q' +\frac{1}{r}) +
{\vp}'q' -{q'}^2 s - V\f0 {q'}^2 =0, \nonumber \\
m'\frac{2}{r}+n'(q'+\frac{2}{r})-{\vp}'q'-2\frac{l-m}{r^2}+{q'}^2 s +
V\f0 {q'}^2 =0.
\end{eqnarray}
Here $q'=\frac{\f0'}{\f0}$ and $ s =l-\vp -2m $.
The equation for the dilaton (\ref{E14}) in the  $O(\e)$ order is:
\begin{equation}   
\label{A5}
{\vp}''+\frac{2}{r}{\vp}'+\frac{1}{2}(2m'+n'-l')q' +{q'}^2 s +\frac{V}{2}
\f0 {q'}^2=0.
\end{equation}

We look for a solution such that $l=-n$, because in this case, as at the
tree level, the components of the metric satisfy the relation $-g_{tt} =
 g^{-1}_{rr}$.

 The system of the equations of motion is consistent with 
the ansatz and we obtain  the following solution
\begin{eqnarray}
m= -n = -A_1\frac{P}{r}  + A_2 \frac{P}{r+P}, \nonumber \\ 
\varphi=\frac{P}{r}(A_1 +\frac{V}{2})+A_2 \frac{P}{r+P},
\label{E26}
\end{eqnarray}
where $A_{1,2}$ are arbitrary constants.
The form of solution is  fixed by the
requirement that at spatial infinity the metric is Lorentzian
and dilaton is unity.

For the metric of the form (\ref{E12}) the expression for the ADM mass is
(for example, \cite{lu})
\begin{equation}
\label{k1}
M= -\left[2(e^{\mu -2\ln r})'r^2 - 2(e^{\l} -e^{\mu-2\ln r} )r \right]
|_{r\rightarrow\infty}
\end{equation}

Substituting expressions (\ref{E25}), where $m$ and $n$ are
 the solutions (\ref{E26}), we obtain the loop-corrected ADM mass
$$
M = 2P(1+\e (A_1 -A_2 )).
$$ 
Requiring that that the ADM mass is equal to the  charge, we have
$$A_1 -A_2 =0 $$
 As yet,
this is an ad hoc requirement, since we did not discuss the issue of the
supersymmetry of solution. For a supersymmetric solution, this means that
solution is BPS extremal. 

The above expressions were
obtained by making two expansions: (i) in $\e$, assuming that corrections
are smaller than the leading-order expressions:
 $|\ln f_0 |>|\e n |,\,|\e l |$, etc., and (ii) by expanding the exponents
$e^{\e n},\,e^{\e l}$, etc. to the first order in $\e$ assuming that
$1>|\e n |,\,|\e l |, \ldots$. The bounds that define the range of validity
of our solution result both in  constraints on the free parameters and
on the domain of variation of $r$.  At small $r$, $ |\ln f_0 |\simeq
\ln\frac{P}{r}$. In this region  we obtain the domain  of validity of
solution
$$1> \frac{\e P V}{r}.$$

At small $r$, the function $\n$ is
\begin{equation}
\label{D1}
\n = \ln\frac{r}{r+P} + \e A \left(\frac{P}{r}-\frac{P}{r+P}\right),
\end{equation}
where $A=A_1 =A_2$.
This expression can be rewritten as the $O(\e)$ term in the expansion of the
function
$$ \n = -\left(\ln\f0 +\e A\frac{{\f0}'}{\f0}\right)$$
which with the same accuracy is the function $\f0$, but with the
shifted argument 
$$\n =\ln\f0 (r+\e A).$$
As it is shown elsewhere, supersymmetric solution is obtained for
$A=\frac{V}{4}$, where $V$ is the Green-Schwarz function.
Extrapolating the metric to  small $r< \e\frac{V}{4}$, we
obtain that the $g_{00}$ component of the metric is
\begin{equation}
\label{D2}
-g_{00} = 1+\frac{P}{r+\e\frac{V}{4}}.
\end{equation} 
Because of quantum corrections, the singularity at $r=0$ is smeared. The
crucial fact is that  the Green-Schwarz $V$ is known to be positive \cite{kou}.

{\large\bf 6.} Another way to obtain the loop-corrected expressions 
for the metric and
moduli is to solve the spinor Killing equations for the $N=2$ supersymmetric 
loop-corrected effective action. In this case we obtain a family of
supersymmetric solutions which is contained in solution (\ref{E26}).  
Details of this work will be reported elsewhere.
 
The problem of string-loop corrections to the classical charged black holes
in the effective $N=2$ supergravity was studied also in papers
\cite{bgl}.
However, the two approaches are rather different. In the cited papers, the loop
corrections were calculated under an assumption that there exists a "small"
modulus and it is possible to expand the prepotential with respect to the ratios
of small and large moduli. For the magnetic black hole, in the universal
sector of the theory, the small modulus is the dilaton, but this modulus
does not enter the loop correction to the prepotential. The remaining two
moduli connected with the metric of the compact two-torus may have
parametric smallness, but not the functional one connected with dependence
on $r$. Properties of the loop-corrected solution depend crucially on the loop
corrections to the gauge couplings. 

 The string-tree-level chiral null model provides a  solution to
 the low energy effective action  which in a special
renormalization  scheme  receives no
$\a'$ corrections \cite{howe,bh2}. The loop-corrected solution 
 no longer is  expressed in terms of  harmonic functions, and the
$\a'$-corrections are present.
However, still it is possible that the $\a'$ corrections are small and can 
be treated perturbatively.
It may be noted here that smearing of the singularity of the point-like
source by $\alpha'$-corrections was discussed previously in \cite{tssm}

I would like to thank L. Pando Zayas with whom this work was initiated, 
R. Kallosh for helpful
correspondence, O. Kechkin and I.Tyutin for remarks and
discussion.
This work was  partially supported by the RFFR  grant No 00-02-17679.

\end{document}